\documentclass[twocolumn]{aastex701}
\usepackage{multirow}

\begin{document}

\title{Search for Radio Pulsations from Neutron Star Candidates in Detached Binaries}

\correspondingauthor{Shi-Jie Gao \& Xiang-Dong Li}

\author[orcid=0000-0002-0822-0337,sname='Gao']{Shi-Jie Gao}
\affiliation{School of Astronomy and Space Science, Nanjing University, Nanjing, 210023, People's Republic of China}
\affiliation{Key Laboratory of Modern Astronomy and Astrophysics, Nanjing University, Ministry of Education, Nanjing, 210023, People's Republic of China}
\email[show]{gaosj@nju.edu.cn}  

\author[orcid=0000-0002-0584-8145,sname='Li']{Xiang-Dong Li}
\affiliation{School of Astronomy and Space Science, Nanjing University, Nanjing, 210023, People's Republic of China}
\affiliation{Key Laboratory of Modern Astronomy and Astrophysics, Nanjing University, Ministry of Education, Nanjing, 210023, People's Republic of China}
\email[show]{lixd@nju.edu.cn}  

\author[orcid=0000-0003-3116-5038,sname='Wang']{Song Wang}
\affiliation{National Astronomical Observatories, Chinese Academy of Sciences, Beijing 100101, People's Republic of China}
\affiliation{Institute for Frontiers in Astronomy and Astrophysics, Beijing Normal University, Beijing, 102206, People's Republic of China}
\email[hide]{songw@bao.ac.cn}
\author[orcid=0000-0002-6871-1752,sname='El-Badry']{Kareem El-Badry}
\affiliation{Department of Astronomy, California Institute of Technology, 1200 East California Boulevard, Pasadena, CA 91125, USA}
\email[hide]{kelbadry@caltech.edu}

\author[orcid=0000-0002-6423-6106,sname='Zhou']{De-Jiang Zhou}
\affiliation{National Astronomical Observatories, Chinese Academy of Sciences, Beijing, 100101, People's Republic of China}
\email[hide]{djzhou@nao.cas.cn}

\author[orcid=0000-0001-5684-0103,sname='Shao']{Yi-Xuan Shao}
\affiliation{School of Astronomy and Space Science, Nanjing University, Nanjing, 210023, People's Republic of China}
\affiliation{Key Laboratory of Modern Astronomy and Astrophysics, Nanjing University, Ministry of Education, Nanjing, 210023, People's Republic of China}
\email[hide]{yixuan@smail.nju.edu.cn}

\author[orcid=0000-0002-9322-9319,sname='Yan']{Zhen Yan}
\affiliation{Shanghai Astronomical Observatory, Chinese Academy of Sciences, Shanghai, 200030, People's Republic of China}
\affiliation{School of Astronomy and Space Science, University of Chinese Academy of Sciences, Beijing, 100049, People's Republic of China}
\email[hide]{yanzhen@shao.ac.cn}

\author[orcid=0000-0002-3386-7159,sname='Wang']{Pei Wang}
\affiliation{National Astronomical Observatories, Chinese Academy of Sciences, Beijing 100101, People's Republic of China}
\affiliation{Institute for Frontiers in Astronomy and Astrophysics, Beijing Normal University, Beijing, 102206, People's Republic of China}
\email[hide]{wangpei@nao.cas.cn}

\author[orcid=0000-0002-5683-822X,sname='Zhou']{Ping Zhou}
\affiliation{School of Astronomy and Space Science, Nanjing University, Nanjing, 210023, People's Republic of China}
\affiliation{Key Laboratory of Modern Astronomy and Astrophysics, Nanjing University, Ministry of Education, Nanjing, 210023, People's Republic of China}
\email[hide]{pingzhou@nju.edu.cn}

\author[orcid=0000-0002-9274-3092,sname='Han']{Jin-Lin Han}
\affiliation{National Astronomical Observatories, Chinese Academy of Sciences, Beijing 100101, People's Republic of China}
\affiliation{School of Astronomy and Space Science, University of Chinese Academy of Sciences, Beijing, 100049, People's Republic of China}
\email[hide]{hjl@nao.cas.cn}

\begin{abstract}
Recent optical astrometric and spectroscopic surveys have identified numerous neutron star (NS) candidates in non-accreting detached binary systems, but their compact-object nature remains unconfirmed. In this work, we present targeted radio observations of 31 such candidates using the Five-hundred-meter Aperture Spherical radio Telescope (FAST), the Robert C. Byrd Green Bank Telescope, and the Shanghai TianMa Radio Telescope. Over a total of 46.65 hours of observing time, we detected neither periodic nor single-pulse radio emissions. {These nondetections place stringent upper limits on the flux densities of any potential radio signals, reaching $ \sim4~\mu\rm Jy$ for periodic emission and $\sim 10~{\rm mJy}$ for single pulses with FAST}. {Since our observations are highly sensitive and the flux density upper limits are well below the median fluxes of known Galactic pulsars, this suggests that geometric beaming is the most likely explanation for the non-detections if these objects are indeed pulsars.} {Alternatively, the NSs may be sufficiently old ($\gtrsim 10~{\rm Gyr}$) and have become intrinsically radio-quiet. In this case,} our findings highlight the inherent difficulty of confirming NSs in such old detached binary systems through radio pulsation searches.

\end{abstract}


\section{Introduction} 

\begin{figure*}
    \centering
    \includegraphics[width=\linewidth]{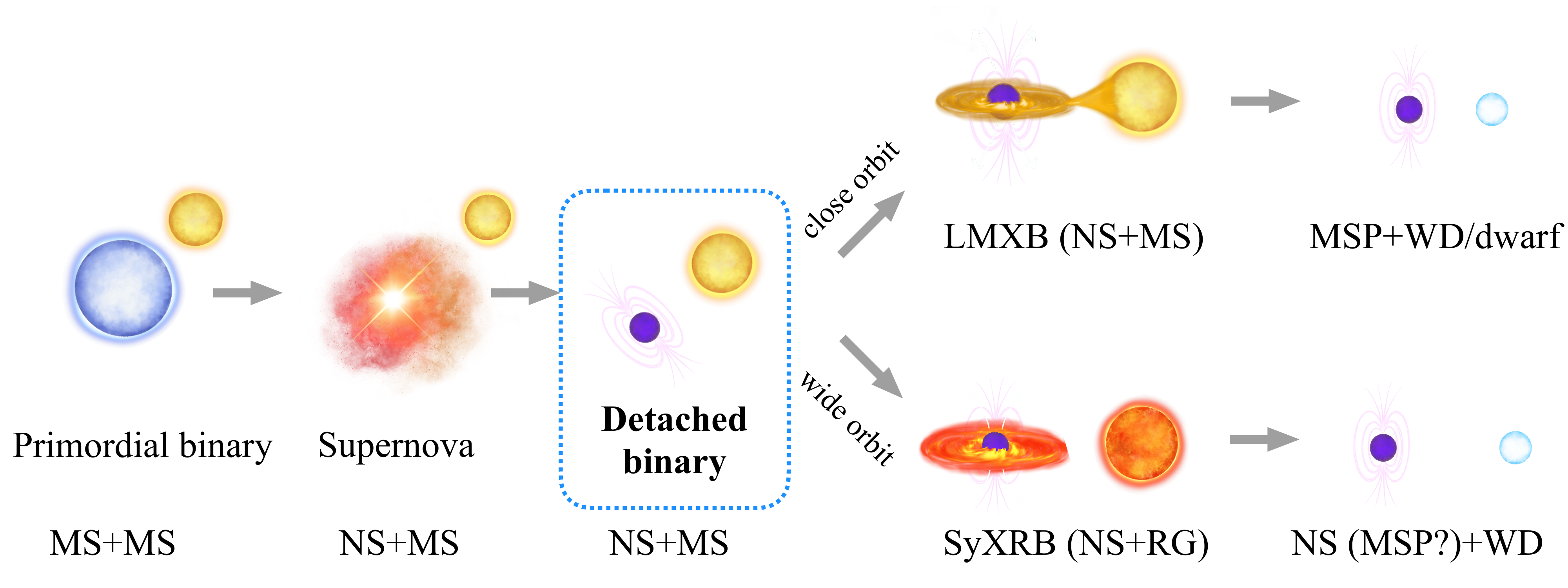}
    \caption{Formation and subsequent evolution of detached binaries with an NS and a low-mass star {\citep[see also][]{Bhattacharya+1991,Tauris+2006,pbse}}. The detached binary stage is highlighted by {the} blue dashed-line box. MS: main-sequence star, NS: neutron star, LMXB: low-mass X-ray binary, MSP: {millisecond} pulsar, WD: white dwarf, SyXRB: symbiotic X-ray binary, RG: red giant.\label{fig:vhd}}
\end{figure*}

The Milky Way is predicted to host $\sim 10^9$ neutron stars (NSs) \citep{Chawla+2022}, yet only a few thousand have been identified to date. This discrepancy poses a significant challenge for characterizing the overall NS population. Most known NSs are detected via their radio pulsations \citep{ATNF}, while others are identified as X-ray sources in accreting binary systems \citep{LMXBcatlog,HMXBcatlog}, isolated X-ray dim NSs \citep{Haberl+2007}, or through associations with supernova remnants \citep{Staelin+1968,Liu+2024}.

In contrast, NSs in detached binaries are paired with an optically luminous, non-degenerate companion star and lack the strong X-ray emission characteristic of accreting systems. Recent optical surveys that combine high-precision astrometry, high-cadence photometry, and radial velocity measurements have proven effective in uncovering such systems \citep{Gaia+2023,Mu+2022}. Multiple research teams have leveraged data from surveys like Gaia, LAMOST, and others to identify promising NS candidates in binaries \citep[e.g.,][]{Yi+2022,GaiaNS1,El-Badry+2024}.

Studying NSs in detached binaries provides critical insights into supernova mechanisms in binary systems and the formation pathways of NS X-ray binaries. \autoref{fig:vhd} illustrates the formation and evolutionary pathways of a detached binary comprising an NS and a low-mass companion ($\lesssim1.5~{M_\odot}$). If the supernova does not disrupt the binary, the nascent NS typically emerges as a radio pulsar in a detached binary system. However, these young NSs are only detectable as radio pulsars for $\sim 10^7~{\rm yr}$ due to spin-down and {and possibly also magnetic field decay} \citep{hpa}. As the binary evolves, mass transfer may occur via Roche-lobe overflow or stellar wind accretion, depending on the orbital separation{ and the mass and evolutionary stage of the companion star} \citep{Podsiadlowski+2002,Shao+2015}. In tight binaries, mass transfer can recycle the old NS into a millisecond pulsar (MSP), typically with a helium white dwarf or a low-mass dwarf-like companion \citep{Bhattacharya+1991,Tauris+2006}. In contrast, wider binaries may evolve into symbiotic X-ray binaries through wind accretion from a low-mass red giant companion \citep{Mikoajewska+2012,Yungelson+2019,Nagarajan+2024}.

Unlike those in low-mass X-ray binaries, NSs in detached binaries evolve similarly to isolated pulsars, as they undergo no significant binary interactions and therefore exhibit little or no X-ray emission. Consequently, detecting radio pulsations remains the most direct evidence for the presence of an NS companion in such systems. Given that the companions in these NS candidate systems are low-mass main-sequence stars ($\lesssim 1~{M_\odot}$) and the NS formation delay time is negligible, the binaries likely share the same age as the NSs, typically $\gtrsim 10^{10}~{\rm yr}$. While isolated pulsars of such ages are expected to fall below the pulsar death line and cease radio emission \citep{Chen+1993,Zhang+2000}, both detections and nondetections of radio pulsations are critical for understanding pulsar radio emission mechanisms and NS formation in binaries.

Motivated by these considerations, we conducted targeted radio observations of 31 NS candidates in detached binaries using state-of-the-art radio telescopes. The rest of this paper is organized as follows: {Section}~\ref{sec:obs} describes the target selection, observational details, and data reduction procedures; Section~\ref{sec:res} presents the results and discussion; and Section~\ref{sec:sum} concludes with a summary of this work.

\section{observations and data reduction}\label{sec:obs}
\subsection{Target Selection and Observations}
\begin{table*}
\centering
    \caption{NS candidates observed with radio telescopes in this work. Their right ascension (R.A.), declination (Decl.), orbital period $P_{\rm orb}$, masses of the luminous optical stars ($M_{\rm O}$) and NS candidates ($M_{\rm NS}$), eccentricities ($e$) and distances ($d$) are listed. The dispersion measure DM$_{d}$ for each targets estimated using models YMW16 and NE2001 are listed, as well as the maximum value DM$_{\rm max}$ in the directions of the targets. Sources without measured eccentricities are assumed to have circular orbits ($e = 0$) during radial velocity fitting as adopted in their respective references.}
    \begin{tabular}{ccc|ccccc|cccc}
        \hline
        Targets&R.A. & Decl.& $P_{\rm orb}$ & $M_{\rm O}$ & $M_{\rm NS}$ & $e$  & $d$ & \multicolumn{2}{c}{YMW16}&\multicolumn{2}{c}{NE2001}\\
        &&&&&&&&DM$_{d}$&DM$_{\rm max}$&DM$_{d}$&DM$_{\rm max}$\\
        &($^{\rm h}:^{\rm m}:^{\rm s}$)&($^{\circ}:':''$)& $(\rm d)$& $(M_\odot)$& $(M_\odot)$&& $(\rm kpc)$ & \multicolumn{4}{c}{$(\rm pc~cm^{-3})$}\\
        \hline
        J0036$-$0932$^{\rm a}$&00:36:11.78&   $-$09:32:38.1& 719.8 & 0.94  & 1.36    & 0.40 & 0.60 & 6.9  & 20.5  & 15.2 & 31.4  \\
        J0230+5950$^{\rm a}$ &02:30:07.56&   +59:50:47.4& 1029  & 1.11  & 1.40    & 0.75 & 0.40 & 7.1  & 296.0 & 4.5  & 212.9 \\
        J0334+0009$^{\rm b}$&03:34:55.36&    +00:09:09.9&510.7&1&2.75&0.28&0.29&16.9&44.6&4.7&41.6\\
        J0553$-$1349$^{\rm a}$&05:53:53.16&   $-$13:49:55.2& 189.1 & 0.98  & 1.33    & 0.39 & 0.40 & 19.8 & 98.7  & 5.9  & 71.6  \\
        J0616+2319$^{\rm c}$ &06:16:35.53&   +23:19:09.4& 0.867 & 1.7   & 1.2     &$-$& 1.11 & 50.8 & 269.3 & 40.9 & 184.9\\
        J0634+6256$^{\rm a}$ &06:34:44.56&   +62:56:01.7& 1046  & 1.18  & 1.48    & 0.56 & 1.45 & 28.2 & 77.0  & 36.2 & 69.9  \\
        J0705$-$0323$^{\rm d}$&07:05:30.23&   $-$03:23:32.2& 27.3  & $-$    & $>1.07$ &$-$& 1.04 & 34.2 & 280.9 & 16.0 & 190.5 \\
        J0724+4040$^{\rm d}$ &07:24:41.50&	+40:40:13.2& 8.2   & $-$    & $>1.47$ &$-$& 1.48 & 40.3 & 81.5  & 41.2 & 70.7  \\
        J0744+3954$^{\rm e}$ &07:44:32.30&   +39:54:21.5& 0.452 & 0.74  & $>0.73$ &$-$& 0.49 & 21.1 & 69.2  & 13.7 & 63.1  \\
        J0824+5254$^{\rm a}$ &08:24:36.94&   +52:54:27.3& 1027  & 1.10  & 1.60    & 0.69 & 0.61 & 11.8 & 42.2  & 15.4 & 48.6  \\
        J0853+1320$^{\rm f}$ &08:53:04.57&   +13:20:32.2& 0.79  & 0.83  & $>0.76$ &$-$& 0.79 & 15.8 & 49.1  & 4.8  & 50.6\\
        J1007+3408$^{\rm g}$ &10:07:21.57&   +34:08:23.7& 999   & 0.6   & 1.2     &$-$& 0.64 & 6.4  & 22.9  & 16.8 & 36.2  \\
        J1007+4453$^{\rm g}$ &10:07:08.81&   +44:53:47.8& 922   & 0.65  & 1.29    &$-$& 0.70 & 7.2  & 23.7  & 17.9 & 36.3   \\
        J1046+1002$^{\rm g}$ &10:46:05.96&   +10:02:58.3& 570   & 0.85  & 1.35    &$-$& 0.55 & 6.9  & 23.9  & 13.2 & 34.9  \\
        J1048+6547$^{\rm a}$ &10:48:59.44&   +65:47:55.6& 814.2 & 1.00  & 1.68    & 0.38 & 1.11 & 13.1 & 28.0  & 24.0 & 37.0  \\
        J1116+5543$^{\rm h}$&11:16:29.94&   +55:43:42.2& 46.1  &0.28   &1.48     &0.02   &0.86&9.5&23.2&19.4&33.2\\
        J1150$-$2203$^{\rm a}$&11:50:52.82&   $-$22:03:50.5& 631.8 & 1.18  & 1.39    & 0.55 & 0.58 & 19.8 & 44.9  & 13.2 & 48.0  \\
        J1205+6914$^{\rm g}$ &12:05:24.72&   +69:14:27.8& 796   & 1.10  & 1.36    &$-$& 1.20 & 14.1 & 27.9  & 24.6 & 36.1  \\
        J1208+3111$^{\rm e}$ &12:08:02.64&   +31:11:03.9& 0.463 & 0.74  & $>0.66$ &$-$& 0.09 & 1.0  & 18.8  & 0.5  & 19.9  \\
        J1432$-$1021$^{\rm a}$&14:32:20.69&   $-$10:21:58.9& 730.9 & 0.79  & 1.90    & 0.12 & 0.73 & 10.6 & 30.9  & 10.6 & 36.5  \\
        J1449+6919$^{\rm a}$ &14:49:18.28&   +69:19:02.3& 632.7 & 0.91  & 1.26    & 0.26 & 0.55 & 7.5  & 29.5  & 8.0  & 36.5  \\
        J1503+2243$^{\rm e}$ &15:03:35.90&   +22:43:22.7& 1.564 & 0.86  & $>0.73$ &$-$& 0.70 & 8.9  & 23.3  & 5.1  & 22.9  \\
        J1622+1647$^{\rm g}$ &16:22:31.74&   +16:47:48.2& 777   & 0.90  & 1.06    & $-$& 0.63 & 8.8  & 32.6  & 4.2  & 37.5   \\
        J1733+5808$^{\rm a}$ &17:33:34.88&   +58:08:46.5& 570.9 & 1.16  & 1.36    & 0.31 & 0.69 & 9.7  & 38.1  & 7.5  & 44.5  \\
        J1739+4502$^{\rm a}$ &17:39:56.06&  +45:02:17.3& 657.4 & 0.78  & 1.38    & 0.68 & 0.89 & 11.9 & 39.1  & 10.5 & 47.4  \\
        J1812+2409$^{\rm g}$ &18:12:15.10&   +24:09:14.2& 728   & 0.80  & 1.18    &  $-$ & 1.08 & 16.3 & 67.8  & 12.1 & 83.2    \\
        J1822+0400$^{\rm d}$ &18:22:29.84&   +04:00:20.8& 11.96 & $-$    & $>0.71$ &$-$ & 0.36 & 5.4  & 177.0 & 3.4  & 253.9 \\
        J1934+0750$^{\rm d}$ &19:34:46.36&   +07:50:30.5& 10.83 & $-$    & $>0.91$ &  $-$& 1.34 & 31.4 & 196.7 & 15.2 & 301.0 \\
        J1949+0129$^{\rm g}$ &19:49:42.67&   +01:29:31.1& 691   & 0.75  & 1.23    & $-$ & 2.05 & 9.8  & 112.9 & 6.6  & 149.6 \\
        J2102+3703$^{\rm a}$ &21:02:28.72&   +37:03:49.3& 480.9 & 1.03  & 1.44    & 0.43 & 0.66 & 10.0 & 165.6 & 7.3  & 202.5 \\
        J2145+2837$^{\rm a}$ &21:45:24.78&   +28:37:25.5& 889.8 & 0.95  & 1.40    & 0.58 & 0.24 & 3.3  & 61.6  & 2.4  & 72.8  \\
        \hline
    \end{tabular}
    \tablerefs{a, \cite{El-Badry+2024}; b, \cite{Shahaf+2023}; c, \cite{Yuan+2022}; d, \cite{Jayasinghe+2023}; e, \cite{Mu+2022}; f, \cite{Li+2022}; g, \cite{Ganguly+2022} and h, \cite{Zhao+2024}.}
    \label{tab:info}
\end{table*}

The Gaia mission has monitored over $2\times10^{9}$ stars across the sky, providing astrometry for $10^5$ binaries and both astrometric and radial velocity solutions for $\sim 3\times10^4$ binaries \citep{Gaia+2023,El-Badry+2024R}. Meanwhile, large-scale spectroscopic surveys such as LAMOST have collected multi-epoch spectra of millions of stars, enabling the detection of unseen companions through periodic Doppler shifts in the luminous optical star's spectrum \citep{Cui+2012,Mu+2022}. By cross-matching these astrometric and spectroscopic datasets, researchers have identified dozens of binaries whose dynamical mass estimates and orbital parameters classify them as hidden non-accreting NS candidates \citep[e.g.,][]{Yi+2022,Yuan+2022,Li+2022,Ganguly+2022,Jayasinghe+2023,Shahaf+2023,GaiaNS1,El-Badry+2024,Zhao+2024b,Zhao+2024}. While most of them are X-ray and UV faint, often falling below current telescope sensitivities \cite[e.g.,][]{Sbarufatti+2024}, deep radio pulsation searches can decisively confirm the NS nature of these systems.

We selected 31 NS candidates from the literature for radio follow-up observation based on their estimated masses and sky positions. Their parameters and references are listed in \autoref{tab:info}. The selection criteria were as follows: (1) binaries were required to have either the NS candidate’s mass or the mass function $\lesssim 2.5~{M_\odot}$ to exclude the possibility of the compact object being a black hole; (2) binaries exhibiting near-ultraviolet excesses were excluded, as such {excesses indicate a higher} probability of containing white dwarfs rather than NSs \citep[see][]{Ganguly+2022}; and (3) binaries located outside the sky coverage of the available radio telescopes were omitted.

\begin{table*}
    \centering
    \caption{Parameters of radio telescopes used in this study. The columns list radio frequency band, system temperature $T_{\rm sys}$, telescope gain $G$, system equivalent flux density (SEFD), central frequency $f_{\rm c}$, total bandwidth $\Delta f$ and sampling time $\Delta t$.\label{tab:tel}}
    \begin{tabular}{ccccccccc}
    \hline
    Telescope&Band&$T_{\rm sys}$&$G$&SEFD&$f_{\rm c}$&$\Delta f$&$\Delta t$&Ref.\\
    &&(K)&($\rm K\,Jy^{-1}$)&Jy&(GHz)&(MHz)&($\mu$s)&\\
    \hline
    FAST&$L$-band&20&16&1.25&1.25&500&49.152&\cite{Jiang+2020}\\
    GBT&$L$-band&24&2&12&1.5&800&81.92&\cite{Prestage+2009}\\
    TMRT&$C$-band (bank~1)&$-$&$-$&30&4.82&1000&65.536&\cite{Yan+2024}\\
    TMRT&$C$-band (bank~2)&$-$&$-$&30&7.0&1000&65.536&\cite{Yan+2024}\\
    \hline
    \end{tabular}
\end{table*}

Radio pulsation searches were performed using three facilities: the Five-hundred-meter Aperture Spherical radio Telescope \citep[FAST,][]{Nan+2008,Nan+2011}, the Robert C. Byrd Green Bank Telescope \citep[GBT,][]{Prestage+2009}, and the Shanghai TianMa Radio Telescope \citep[TMRT,][]{Yan+2017,Yan+2024}. \autoref{tab:tel} lists the telescope parameters in this study, including the system temperature $T_{\rm sys}$, telescope gain $G$, system equivalent flux density (SEFD, $T_{\rm sys}/G$), central frequency $f_{\rm c}$, total bandwidth $\Delta f$ and sampling time $\Delta t$.

FAST observations\footnote{Project IDs: PT2023\_0167 (PI: Shi-Jie Gao), PT2024\_0015 (PI: Shi-Jie Gao) and PT2024\_0197 (PI: Song Wang).} were carried out using the center beam of the $L$-band 19-beam receiver at a central frequency of 1.25~GHz and a bandwidth of 500~MHz \citep{LiDi+2018,Jiang+2020}. The data were sampled at a time resolution of $49.152~{\rm \mu s}$ and recorded in PSRFITS-format \citep{psrfits} files with 4096 (2048 channels for J0334+0009 and J1116+5543) frequency channels and four polarizations {parameters}.

GBT observations\footnote{Project ID: GBT24A\_092, PI: Shi-Jie Gao.} utilized the $L$-band receiver with the Versatile GBT Astronomical Spectrometer's \cite[VEGAS,][]{vegas} Pulsar Mode (VPM). The observations were conducted at a central frequency of 1.5~GHz with an 800~MHz bandwidth. A notch filter was implemented to suppress radio frequency interference (RFI) from a nearby Air Surveillance Radar between 1.2 and 1.34~GHz. Data were acquired at a time resolution of $81.92~{\mu \rm s}$ across 4096 frequency channels and the total intensity from two summed polarizations was recorded. In each observation session, a nearby known bright pulsar was observed to verify the proper functioning of the telescope.

TMRT observations\footnote{Project ID: TMP035, PI Shi-Jie Gao.} were conducted with the $C$-band receiver and its digital backend system (DIBAS), which is based upon the design of VEGAS {and provides similar capabilities to the} GBT Ultimate Pulsar Processing Instrument \citep[GUPPI,][]{guppi}. Observations were conducted at central frequencies of 4.82~GHz and 7.0~GHz, with each band offering a 1000~MHz bandwidth divided into 512 channels. Specifically, J1007+3408, J1007+4453 and J1046+1002 were observed at 4.82~GHz, while the remaining targets {were observed simultaneously at} both frequency bands. Data were recorded in full Stokes with a sample time of $65.536~{\mu \rm s}$. Prior to the main observations, the bright pulsar PSR~B1133+16 was observed to verify system performance.

\subsection{Periodic Signal Searches}
\begin{figure*}
    \centering
    \includegraphics[width=\linewidth]{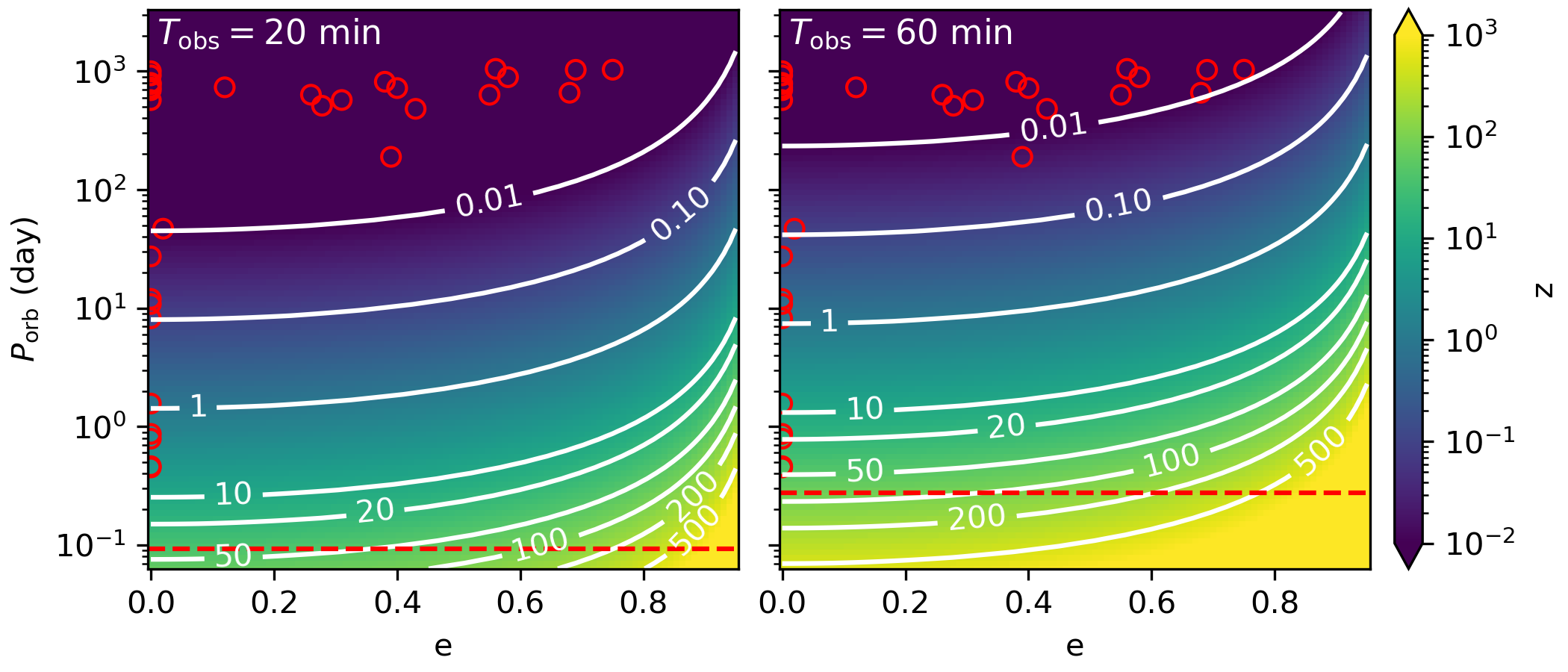}
    \caption{Fourier frequency derivative $z$ required to account for orbital acceleration effects in pulsar searches in detached binaries. $z$ values are calculated for a binary system consisting of a $1.4~{M_\odot}$ NS and a $1.0~{M_\odot}$ companion in an orbit observed at a true anomaly of $A_{\rm T}=-90~\deg$ in an edge-on configuration ($i=90~\deg$) with an argument of periapsis $\omega=0~\deg$. The left and right panels correspond to integration times of $T_{\rm obs}=20~{\rm min}$ and $60~{\rm min}$. Spin frequency of the pulsar is assumed to be $1~{\rm Hz}$ and the harmonic summing set to $h=16$. Red circles mark NS candidates with known orbital periods and eccentricities (see \autoref{tab:info}; with $e=0$ assumed when unspecified). Red dashed lines indicate the $T_{\rm obs}<0.15P_{\rm orb}$ condition, below which an orbital jerk search is necessary.\label{fig:zmax}}
\end{figure*}

For data preprocessing, we employed the \texttt{filtool} command from \texttt{PulsarX} \citep{PulsarX} to mitigate RFI, optimize the bandpass and convert data from PSRFITS-format to filterbank-format \citep{filterbank} data with 1024 frequency channels (for FAST and GBT data) for subsequent pulsation searches.

Periodic pulsation searches were conducted using the PulsaR Exploration and Search TOolkit\footnote{\url{https://github.com/scottransom/presto}} \citep[\texttt{PRESTO},][]{Ransom+2011}. We first used the \texttt{PRESTO} routine \texttt{rfifind} to identify and excise corrupted channels and integrations affected by strong and periodic RFI. Subsequently, the \texttt{PRESTO} command \texttt{prepsubband}\footnote{Utilizing a GPU-version, \url{https://github.com/zdj649150499/Presto_GPU}.} was employed to generate de-dispersed barycentric time-series. \autoref{tab:info} presents the estimated DM values ${{\rm DM}_{d}}$ for each targets based on their positions and distances, as well as the maximum value $\rm DM_{max}$ along their line of sight. These values are derived using two Galactic electron density models: YMW16 \citep{YMW16} and NE2001 \citep{Cordes+2002}. To accommodate uncertainties in unknown DM contributions and the possibility of background pulsars or transients, dedispersion was performed over a DM range of $0-1000~{\rm pc~cm^{-3}}$. {The dedispersion step sizes were selected following the methodology implemented in the \texttt{PRESTO} Python script \texttt{DDplan.py}, which determines DM steps to limit the time smearing between adjacent dispersion measures to a small fraction of the effective time resolution, thereby balancing sensitivity and computational efficiency.}

The orbital periods $P_{\rm orb}$ of these binaries are in the range of $\sim 1-10^3~{\rm day}$, which are significantly longer than the integration times $T_{\rm obs}$ of $\sim$20 min--1 hour. Over a sufficiently small fraction of the orbit ($T_{\rm obs}\lesssim P_{\rm orb}/10$), the pulsar's acceleration can be approximated as constant, resulting in a linear drift of the measured spin frequency $f$ over time \citep{Johnston+1991,Andersen+2018}. A Fourier-domain acceleration search technique \citep{Ransom+2002,Dimoudi+2018} was therefore employed to search for pulsars in binaries. In this method the Fourier frequency derivative is defined as $z=\dot f T_{\rm obs}^2$ (where $\dot f$ is the spin frequency derivative). This parameter $z$ corresponds to a drift in the physical linear acceleration due to orbital motion, given by \citep{Andersen+2018}
\begin{equation}\label{eq:a}
    \alpha=\frac{\dot f c}{hf}=\frac{zc}{hfT_{\rm obs}^2},
\end{equation}
where $c$ is the speed of light in vacuum, $f$ is the spin frequency and $h$ is the harmonic summing factor with $h=1$ representing the fundamental.

For a pulsar in a binary with orbital period $P_{\rm orb}$, semimajor axis $a$, eccentricity $e$, inclination $i$, argument of periapsis $\omega$ and true anomaly $A_{\rm T}$, the line-of-sight acceleration is given by \citep{Freire+2001}
\begin{equation}\label{eq:al}
    \alpha=-\left(\frac{2\pi}{P_{\rm orb}}\right)^2\frac{a\sin i}{(1-e^2)^2}(1+e\cos A_{\rm T})^2\sin(A_{\rm T}+\omega).
\end{equation}
By combining \autoref{eq:a} and \autoref{eq:al}, the necessary $z$ values were computed and are shown in \autoref{fig:zmax}.
Known NS candidates (see \autoref{tab:info}) with measured $P_{\rm orb}$ and $e$ (with $e=0$ assumed when not provided) are indicated by red circles in \autoref{fig:zmax}.
For systems with much tighter orbits, where $T_{\rm obs}>0.15P_{\rm orb}$, the acceleration changes over the observation interval and an orbital jerk search is required to improve sensitivity \citep{Andersen+2018}. The condition $T_{\rm obs}>0.15P_{\rm orb}$ for jerk searches restricts to $P_{\rm orb}<0.093~{\rm day}$ and $<0.28~{\rm day}$ for $T_{\rm obs}=20~{\rm min}$ and $60~{\rm min}$, respectively. {These} restrictions are indicated by the red dashed lines in \autoref{fig:zmax}. Since all targets in this work satisfy {{$T_{\rm obs}\ll 0.15P_{\rm orb}$}}, an acceleration search with a maximum $z$ value of 200 is adopted to sufficiently reveal potential pulsars while balancing sensitivities and computational cost.

The \texttt{PRESTO} routine \texttt{accelsearch} was employed to search for periodic signals, summing up to $h=16$ harmonics with a maximum $z$ value of $z_{\rm max}=200$. Candidates generated by \texttt{accelsearch} were subsequently winnowed using the Python script \texttt{ACCEL\_sift.py} provided with \texttt{PRESTO}, and a sigma threshold of $4.0$ is applied. Finally, folded diagnostic plots were produced using the \texttt{PRESTO} routine \texttt{prepfold} and the resulting candidates were inspected visually to identify the most promising candidates.

Additionally, a phase-coherent search technique based on the fast folding algorithm \citep[FFA,][]{Morello+2020} was employed. Unlike Fourier-transform-based methods, FFA is a time-domain periodicity search method and offers superior sensitivity to
long-period and nulling pulses. The FFA search was implemented using the \texttt{RIPTIDE} package \citep{Morello+2020}. The \texttt{rffa} pipeline within \texttt{RIPTIDE} was applied to the time-series data produced by \texttt{prepsubband} for periodic signals with periods ranging from 0.1 to 30~{\rm s}. Diagnostic plots generated by the \texttt{rffa} pipeline were inspected visually, and promising candidates were subsequently folded with \texttt{prepfold} to assess their astrophysical or RFI origins.

\subsection{Single-pulse Searches}

Single-pulse searches were conducted using both \texttt{TransientX} \footnote{\url{https://github.com/ypmen/TransientX}} \citep{TransientX} and \texttt{Heimdall} \footnote{\url{https://sourceforge.net/projects/Heimdall-astro}} \citep{Barsdell+2024} to provide cross-verification.

\texttt{TransientX} is a high-performance CPU-based single-pulse search software. For pulse searches using \texttt{TransientX}, a dedispersion grid with a DM range of $0-2000~{\rm pc~cm^{-3}}$ with step sizes of $0.1-1~{\rm pc~cm^{-3}}$ was employed, which was generated using the \texttt{PRESTO}'s Python script \texttt{DDplan.py}. The maximum pulse width was set to the default value of $\sim 50~{\rm ms}$ and a signal-to-noise ratio (S/N) threshold of 7.0 was applied. \texttt{Heimdall} is a GPU-accelerated pipeline for radio transient detection. {For pulse searches using} \texttt{Heimdall}, a similar DM range of $0-2000~{\rm pc~cm^{-3}}$ was used, with a dedispersion plan implemented within the software. The maximum boxcar width was set to 1024 samples, corresponding to a maximum pulse width of $\sim 50~{\rm ms}$, $83.9~\rm ms$ and $67~\rm ms$ for FAST, GBT and TMRT observations. Candidates with similar boxcar or DM trials were clustered within \texttt{Heimdall} and a S/N threshold of 7.0 was applied.

Diagnostic plots, including the dedispersed pulse profile, dedispersed frequency versus time, and DM space were generated using the single-pulse analysis software \texttt{YOUR} \citep{Aggarwal+2020}. Given the large number of candidates, we employed \texttt{FETCH} \citep{FETCH}, an open-source machine learning tool utilizing convolutional neural networks, to analyze the diagnostic plots. A candidate was considered promising if at least one of the five trained models classified it favorably. Finally, the filtered diagnostic plots were inspected visually to identify the most promising single-pulse candidates.
\subsection{Pipeline Validation}

To validate our search pipelines, we first analyzed data from PSR~J1503+2111, which was recorded during observations of target J1503+2243 with GBT using similar instrumental settings. PSR~J1503+2111 is a nearby isolated normal pulsar with a spin period of 3.31~s and a DM of $3.26~{\rm pc~cm^{-3}}$ \citep{Bilous+2016}. Details of the pipeline validation and criteria for distinguishing astrophysical signals from RFI are provided in \autoref{sec:app}. Briefly, the Fourier-transform-based acceleration search using \texttt{PRESTO} recovered PSR~J1503+2111 (see \autoref{fig:pfd_psr} in Appendix for more details). {The} FFA pipeline also successfully detected the pulsar (see \autoref{fig:ffa_psr} for more details), indicating its effectiveness in identifying long-period pulsars. Although PSR~J1503+2111 is relatively faint, with an estimated peak flux of $<~50~\rm mJy$ (based on its 149 MHz peak flux density and a spectral index of $-1.2$, \citealt{Bilous+2016}), our single-pulse search pipelines successfully detected several weak single pulses (see an example in \autoref{fig:sp_psr}). The successful detection of both periodic radio pulses and faint single pulses confirms the reliability and sensitivity of our search methods.

\section{results and discussion}\label{sec:res}

After careful inspection of the search results, we found no compelling evidence for either periodic or single-pulse emission from any of the NS candidates. In addition, the FAST's data were independently processed using the pulsar search pipelines employed in the FAST Galactic Plane Pulsar Snapshot (GPPS) survey \citep{GPPSI,GPPSII,GPPSVI}. Both the periodic search (without acceleration search, technical details in \citealt{GPPSI}) and the single-pulse search (technical details in \citealt{GPPSII}) also yielded no detections. We next discuss the implications of nondetections, focusing on search sensitivity, detection probability, whether the NSs remain active as radio pulsars and comparison with known binary pulsars with optical companions.

\subsection{Search Sensitivities}

For periodic signals, we use the radiometer equation to estimate the upper limit of the flux density $S_{\rm min}$ for pulsar candidates \citep{hpa}, i.e.,
\begin{equation}
    S_{\rm min}=\frac{{\rm S/N}_{\rm min} T_{\rm sys}}{G\sqrt{N_{\rm p} \Delta f_{\rm eff} T_{\rm obs}}}\sqrt{\frac{W_{\rm eff}}{P-W_{\rm eff}}}, 
\end{equation}
where ${\rm S/N}_{\rm min}$ is the signal-to-noise ratio threshold for the periodic search, $N_{\rm p}=2$ is the number of summed polarizations in our observations, and $\Delta f_{\rm eff}$ is the effective bandwidth after removing bad edges and RFI which is $\sim 75\%$, $\sim 75\%$, $\sim 50\%$ and $\sim 80\%$ the total bandwidth $\Delta f$ for FAST, GBT, TMRT (4.82~GHz) and TMRT (7.0~GHz) observations, respectively. $T_{\rm obs}$ is the total observation duration, which is listed in \autoref{tab:obs}. The parameters $P$ and $W_{\rm eff}$ represent the pulsar period and the effective pulse width, respectively. The effective pulse width is given by \begin{equation}
    W_{\rm eff}=\sqrt{W_{\rm i}^2+\Delta t^2+\tau_{\rm scatter}^2+\Delta t_{\rm DM}^2},
\end{equation}
where $W_{\rm i}$ is the intrinsic pulse width, $\tau_{\rm scatter}$ is the scattering timescale, and $\Delta t_{\rm DM}$ is the uncorrected dispersion delay within a frequency channel.
The pulsars in these NS candidates are likely to be normal pulsars rather than recycled MSPs, and their DM values are not expected to be significantly large (see \autoref{tab:info}). Both the scattering timescale $\tau_{\rm scatter}$ ($<2~{\mu \rm s}$ at $1.4~{\rm GHz}$ assuming DM=100$~{\rm pc~cm^{-3}}$, \citealt{Krishnakumar+2015}) and the uncorrected dispersion delay $\Delta t_{\rm DM}$ are expected to be negligible.
Consequently, the effective pulse width is primarily determined by the intrinsic pulse width and an effective pulse width ratio $W_{\rm eff}/P\simeq 0.1$ is assumed.

The upper limits on radio flux density $S_{\rm min}$ for all NS candidates, derived using a minimum signal-to-noise ratio ${\rm S/N}_{\rm min}=10$ are summarized in \autoref{tab:obs}. 
The $S_{\rm min}$ values achieved for different telescopes and observation durations are as follows:
for FAST observations, $S_{\rm min}$ can reach as low as $3.71$ and $4.63~{\mu\rm Jy}$, corresponding to $T_{\rm obs}=30$ and $20~{\rm min}$, respectively; for GBT observations, $S_{\rm min}$ can reach $11.34$, $16.04$ and $22.68~{\mu\rm Jy}$, corresponding to $T_{\rm obs}=120$, $60$ and $30~{\rm min}$, respectively; for TMRT observations, $S_{\rm min}=52.70$ and $41.67~{\mu\rm Jy}$ with $T_{\rm obs}=60~{\rm min}$ for $f_{\rm c}=4.82$ and $7.0~{\rm GHz}$, respectively. To facilitate cross-comparison across observing frequencies, the $S_{\rm min}$ values were scaled to a reference frequency $1400~{\rm MHz}$ using a spectral index of $-1.8$ \citep{Maron+2000}. These scaled values are also included in  \autoref{tab:obs}. 

The upper limit on single-pulse flux density ($S_{\rm min}^{\rm SP}$) can be estimated using the radiometer equation
 \citep{Cordes+2003},
\begin{equation}
    S_{\rm min}^{\rm SP}=\frac{S/N_{\rm min}T_{\rm sys}}{G\sqrt{N_{\rm p}\Delta f_{\rm eff}W_{\rm eff}}},
\end{equation}
where ${\rm S/N}_{\rm min}=7$ is the minimum signal-to-noise threshold for detection. The derived $S_{\rm min}^{\rm SP}$ values for our observations are listed in \autoref{tab:obs}, and summarized below assuming an effective pulse width $W_{\rm eff}=1~{\rm ms}$: $10.10$, $63.90$, $210.00$ and $166.02~{\rm mJy}$ for FAST, GBT, TMRT (4.82 GHz) and TMRT (7.0 GHz) observations, respectively. 

\startlongtable
\begin{deluxetable*}{ccccccccc}
\tablewidth{1pt}
\tablecaption{Observation parameters for NS candidates, including UTC time, telescope, central frequency $f_{\rm c}$, integration time $T_{\rm obs}$, the estimated flux density upper limit $S_{\rm min}$ for periodic signals, the 1400 MHz-scaled flux density upper limit $S^{1400}_{\rm min}$, the pseudo-luminosity at 1400~MHz $L^{1400}_{\rm min}$, and the flux density upper limit $S_{\rm min}^{\rm SP}$ for single pulses.\label{tab:obs}}
\tablehead{
         Name& UTC Time & Telescope & $f_{\rm c}$ & $T_{\rm obs}$& $S_{\rm min}$ &$S^{1400}_{\rm min}$&$L^{1400}_{\rm min}$&$S_{\rm min}^{\rm SP}$\\
         &(YYYY-MM-DD hh:mm:ss)&&(GHz)&(min)&($\mu \rm Jy$)&($\mu\rm Jy$)&($10^{-3}\rm mJy~kpc^2$)&(mJy)
}
\startdata
J0036$-$0932& 2024-08-14 19:12:00& FAST &1.25&  20&   4.63&   3.78&   2.27&  10.10\\
J0036$-$0932& 2024-08-30 19:14:00& FAST &1.25&  30&   3.71&   3.03&   1.82&  10.10\\
J0036$-$0932& 2024-05-01 15:20:44& GBT  &1.50&  30&  22.68&  25.68&  15.41&  63.90\\
J0230+5950 & 2024-08-14 20:48:02& FAST &1.25&  20&   4.63&   3.78&   1.51&  10.10\\
J0230+5950 & 2024-08-23 21:39:01& FAST &1.25&  30&   3.71&   3.03&   1.21&  10.10\\
J0230+5950 & 2024-04-30 22:11:16& GBT  &1.50&  60&  16.04&  18.16&   7.26&  63.90\\
J0334+0009 & 2023-08-28 22:03:04& FAST &1.25&  20&   4.63&   3.78&   1.09&  10.10\\
J0553$-$1349& 2024-05-04 21:03:10& GBT  &1.50&  30&  22.68&  25.68&  10.27&  63.90\\
J0616+2319 & 2023-10-22 21:44:55& FAST &1.25&  20&   4.63&   3.78&   4.19&  10.10\\
J0616+2319 & 2024-02-19 12:20:52& FAST &1.25&  80&   2.22&   1.81&   2.01&  10.10\\
J0616+2319 & 2024-02-14 06:10:30& GBT  &1.50& 120&  11.34&  12.84&  14.25&  63.90\\
J0634+6256 & 2024-08-14 01:23:57& FAST &1.25&  20&   4.63&   3.78&   5.47&  10.10\\
J0634+6256 & 2024-08-31 00:18:00& FAST &1.25&  30&   3.71&   3.03&   4.39&  10.10\\
J0634+6256 & 2024-05-01 00:14:49& GBT  &1.50&  55&  16.75&  18.97&  27.50&  63.90\\
J0705$-$0323& 2024-04-11 00:58:36& GBT  &1.50&  60&  16.04&  18.16&  18.88&  63.90\\
J0724+4040 & 2023-10-15 20:39:59& FAST &1.25&  20&   4.63&   3.78&   5.59&  10.10\\
J0744+3954 & 2024-01-05 18:42:02& TMRT &4.82&  60&  52.70& 487.87& 239.06& 210.00\\
J0744+3954 & 2024-01-05 18:42:02& TMRT &7.00&  60&  41.67& 754.98& 369.94& 166.02\\
J0744+3954 & 2024-02-14 08:12:46& GBT  &1.50&  30&  22.68&  25.68&  12.58&  63.90\\
J0824+5254 & 2024-08-15 02:49:03& FAST &1.25&  20&   4.63&   3.78&   2.30&  10.10\\
J0824+5254 & 2024-08-30 02:30:02& FAST &1.25&  30&   3.71&   3.03&   1.85&  10.10\\
J0824+5254 & 2024-05-01 01:16:36& GBT  &1.50&  60&  16.04&  18.16&  11.08&  63.90\\
J0853+1320 & 2023-10-22 22:17:02& FAST &1.25&  20&   4.63&   3.78&   1.25&  10.10\\
J0853+1320 & 2024-03-28 02:50:55& GBT  &1.50&  60&  16.04&  18.16&   5.99&  63.90\\
J1007+3408 & 2024-01-04 20:04:16& TMRT &4.82&  75&  47.14& 436.36& 279.27& 210.00\\
J1007+3408 & 2024-03-05 06:12:40& GBT  &1.50&  60&  16.04&  18.16&  11.62&  63.90\\
J1007+4453 & 2024-01-04 22:48:17& TMRT &4.82&  75&  47.14& 436.36& 305.45& 210.00\\
J1007+4453 & 2024-03-05 05:06:51& GBT  &1.50&  60&  16.04&  18.16&  12.71&  63.90\\
J1046+1002 & 2024-01-04 21:31:14& TMRT &4.82&  75&  47.14& 436.36& 240.00& 210.00\\
J1046+1002 & 2024-03-26 04:38:55& GBT  &1.50&  60&  16.04&  18.16&   9.99&  63.90\\
J1048+6547 & 2024-01-05 16:32:18& TMRT &4.82&  60&  52.70& 487.87& 541.53& 210.00\\
J1048+6547 & 2024-01-05 16:32:18& TMRT &7.00&  60&  41.67& 754.98& 838.03& 166.02\\
J1048+6547 & 2024-03-06 05:23:16& GBT  &1.50&  60&  16.04&  18.16&  20.16&  63.90\\
J1116+5543 & 2023-11-18 01:07:58& FAST &1.25&  20&   4.63&   3.78&   3.24&  10.10\\
J1150$-$2203& 2024-05-01 02:19:58& GBT  &1.50&  55&  16.75&  18.97&  11.00&  63.90\\
J1205+6914 & 2024-03-02 07:11:34& GBT  &1.50&  64&  15.53&  17.58&  21.10&  63.90\\
J1208+3111 & 2023-10-31 01:25:58& FAST &1.25&  20&   4.63&   3.78&   0.34&  10.10\\
J1208+3111 & 2024-03-28 03:58:36& GBT  &1.50&  60&  16.04&  18.16&   1.63&  63.90\\
J1432$-$1021& 2024-09-02 08:17:57& FAST &1.25&  20&   4.63&   3.78&   2.76&  10.10\\
J1432$-$1021& 2024-09-16 06:45:56& FAST &1.25&  30&   3.71&   3.03&   2.21&  10.10\\
J1432$-$1021& 2024-02-16 07:22:13& GBT  &1.50&  50&  17.57&  19.89&  14.52&  63.90\\
J1432$-$1021& 2024-03-26 06:53:16& GBT  &1.50&  60&  16.04&  18.16&  13.26&  63.90\\
J1432$-$1021& 2024-03-28 07:09:24& GBT  &1.50&  50&  17.57&  19.89&  14.52&  63.90\\
J1449+6919 & 2024-01-05 20:51:56& TMRT &4.82&  60&  52.70& 487.87& 268.33& 210.00\\
J1449+6919 & 2024-01-05 20:51:56& TMRT &7.00&  60&  41.67& 754.98& 415.24& 166.02\\
J1449+6919 & 2024-03-06 06:26:12& GBT  &1.50&  65&  15.41&  17.45&   9.60&  63.90\\
J1503+2243 & 2023-11-02 05:28:01& FAST &1.25&  20&   4.63&   3.78&   2.64&  10.10\\
J1503+2243 & 2024-01-05 19:49:17& TMRT &4.82&  60&  52.70& 487.87& 341.51& 210.00\\
J1503+2243 & 2024-01-05 19:49:17& TMRT &7.00&  60&  41.67& 754.98& 528.49& 166.02\\
J1503+2243 & 2024-03-28 05:25:09& GBT  &1.50&  50&  17.57&  19.89&  13.92&  63.90\\
J1622+1647 & 2024-03-07 07:31:52& GBT  &1.50&  55&  16.75&  18.97&  11.95&  63.90\\
J1733+5808 & 2024-08-29 10:33:01& FAST &1.25&  20&   4.63&   3.78&   2.60&  10.10\\
J1733+5808 & 2024-05-05 02:34:48& GBT  &1.50&  60&  16.04&  18.16&  12.53&  63.90\\
J1739+4502 & 2024-05-05 03:31:40& GBT  &1.50&  60&  16.04&  18.16&  16.16&  63.90\\
J1739+4502 & 2024-05-08 06:16:50& GBT  &1.50&  50&  17.57&  19.89&  17.70&  63.90\\
J1812+2409 & 2023-09-29 07:56:03& FAST &1.25&  20&   4.63&   3.78&   4.08&  10.10\\
J1822+0400 & 2023-10-24 10:13:00& FAST &1.25&  20&   4.63&   3.78&   1.36&  10.10\\
J1822+0400 & 2024-04-05 12:16:24& GBT  &1.50&  65&  15.41&  17.45&   6.28&  63.90\\
J1934+0750 & 2023-10-25 10:32:00& FAST &1.25&  20&   4.63&   3.78&   5.06&  10.10\\
J1934+0750 & 2024-04-05 13:23:22& GBT  &1.50&  60&  16.04&  18.16&  24.33&  63.90\\
J1949+0129 & 2024-04-05 14:24:51& GBT  &1.50&  65&  15.41&  17.45&  10.82&  63.90\\
J2102+3703 & 2024-08-15 13:40:56& FAST &1.25&  20&   4.63&   3.78&   2.49&  10.10\\
J2102+3703 & 2024-05-08 07:18:11& GBT  &1.50&  60&  16.04&  18.16&  11.98&  63.90\\
J2145+2837 & 2024-08-30 15:46:56& FAST &1.25&  20&   4.63&   3.78&   0.91&  10.10\\
J2145+2837 & 2024-05-08 08:19:40& GBT  &1.50&  70&  14.85&  16.81&   4.03&  63.90\\
\hline 
\enddata
\end{deluxetable*}

\begin{figure*}[thb!]
    \centering
    \includegraphics[width=0.8\linewidth]{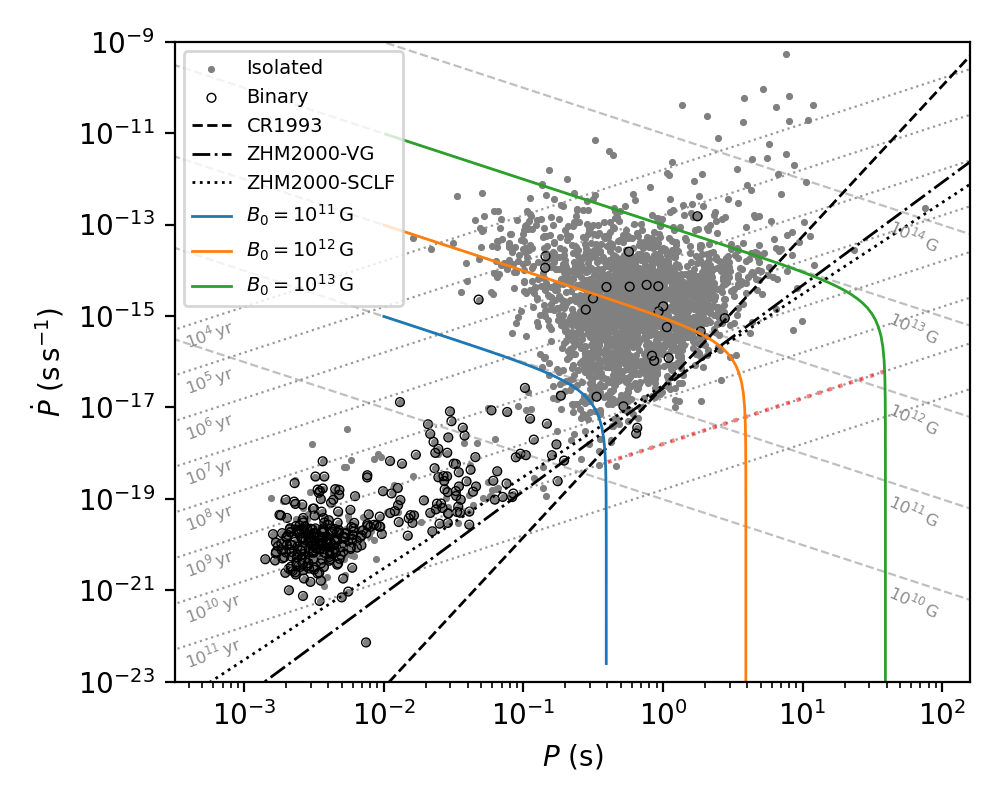}
    \caption{The period--period derivative ($P$--$\dot P$) diagram for known pulsars \citep{ATNF}. Gray dots represent isolated pulsars while black circles denote pulsars in binary systems. The black dashed line correspond to pulsar death line model proposed by \cite{Chen+1993}. The black dash-dotted and dotted lines correspond to pulsar death line models base on curvature radiation from vacuum gap (VG) and space-charged-limited flow (SCLF) \citep{Zhang+2000}. The gray dashed,  and dotted lines represent the lines of constant magnetic field and constant characteristic age \citep{hpa}, respectively. The blue, orange and green curves show spin-down evolutionary tracks with an initial period of 20~ms and $B_{\rm 0}=10^{11},$ $10^{12}$ and $10^{13}~{\rm G}$, respectively.\label{fig:ppdot}}
\end{figure*}

\subsection{Detection Probability}\label{subsec:prob}

If the observed targets are NSs, the absence of detected radio pulsations can be attributed to two primary factors: (1) the pulsar's radio emission beam {does not sweep across the Earth}, and (2) the pulsar is too faint to detect due to significant spin-down, placing it below the pulsar death line and thus lacking the necessary conditions for radio emission.

The luminosity sensitivity limit $L_{\rm min}^{1400}$ of our pulsation search, expressed as the monochromatic luminosities (pseudo-luminosities) at 1400~MHz, is calculated as
\begin{equation}
    L_{\rm min}^{1400}=S_{\rm min}^{1400}d^2,
\end{equation}
where $S_{\rm min}^{1400}$ is the flux density scaled to 1400 MHz and $d$ is the distance to the target. 
The $L_{1400}^{\rm min}$ values for all targets are listed in \autoref{tab:obs}. For targets with multiple observations, only the the most constraining is retained for subsequent calculations.
We adopt a log-normal luminosity function for observed radio pulsars \citep{Szary+2014}
\begin{equation}
    \phi\left(\log \mathcal{L}\right)=\frac{1}{\sqrt{2\pi}\sigma}\exp\left[-\frac{(\log \mathcal{L}-\mu)^2}{2\sigma^2}\right],
\end{equation}
where $\mathcal{L}=L_{\rm 1400}/(\rm mJy~kpc^2)$ is the pseudo-luminosity at 1400~MHz in units of $\rm mJy~kpc^{2}$, and $\mu=0.5$ and $\sigma=1.0$ are the mean and standard deviation of $\log \mathcal{L}$, respectively. The beaming fraction $f_{\rm b}$(\%) depends on the pulsar’s spin period $P$ and is modeled as \citep{Tauris+1998}
\begin{equation}
    f_{\rm b}=9\left[\log\left(\frac{P}{10~{\rm s}}\right)\right]^2+3.
\end{equation}

Accounting for both luminosity and beaming effects, the nondetection probability for the $i$-th target is given by
\begin{equation}
    \mathcal{P}_{{\rm non},i}=1-f_{\rm b}\times\int_{\mathcal{L}_{\rm min},i}^{\infty}\phi(\log \mathcal{L})\mathrm d (\log \mathcal{L}),
\end{equation}
where $\mathcal{L}_{{\rm min},i}$ is the luminosity limit for the $i$-th target. The overall nondetection probability $\mathcal{P}_{{\rm non}}$ is the product of $\mathcal{P}_{{\rm non},i}$ over all targets. Assuming all underlying pulsars have spin periods of $P=0.1$, $1$ and $10~{\rm s}$, the corresponding total nondetection probabilities are $\mathcal{P}_{\rm non}\simeq 2.4\times10^{-7}$, 0.019 and 0.39, respectively. Given the high sensitivity of our observations ($\log \mathcal{L}_{{\rm min},i}\ll\mu$), the nondetection probability is primarily driven by the beaming fraction and is largely insensitive to the specific form of the luminosity function. In this context, the nondetections are most likely attributable to selection effects arising from the geometry of pulsar beaming, if these objects are indeed pulsars.

\subsection{Are the NSs in These Binaries Still Active as Radio Pulsars?}\label{subsec:dl}

In detached binaries, an NS evolves similarly to an isolated pulsar in the absence of significant accretion or interactions with its companion. Under the assumption of pure magnetic dipole radiation, the spin-down rate $\dot P$ of an NS is given by
\begin{equation}
    \dot P=\frac{8\pi^2B^2R^6\sin^2 \alpha}{3c^3IP},
\end{equation}
where $B$ is the surface magnetic field strength, $R$ is the NS radius, $I$ is moment of inertia, and $\alpha$ is the angle between magnetic and rotational axes. Following \cite{Oslowski+2011}, we adopt an exponentially decay model for the magnetic field as a function of time $t$,
\begin{equation}
    B=(B_{\rm 0}-B_{\rm min})\exp(-t/\tau_{\rm B})+B_{\rm min},
\end{equation}
where $B_0$ is the initial magnetic field at birth ($t=0$), $B_{\rm min}$ is the minimum magnetic field, and $\tau$ is the decay timescale. Here we adopt $\sin \alpha=1$, $I=10^{45}~{\rm g~cm^2}$, $R=10^{6}~{\rm cm}$, $B_{\rm min}=10^8~{\rm G}$ and $\tau_{\rm B}=500~{\rm Myr}$.

Using these assumptions, we compute three spin-down evolutionary tracks, each starting with an initial spin period 20~ms and $B_{\rm 0}=10^{11},$ $10^{12}$ and $10^{13}~{\rm G}$. These tracks are plotted in the $P$--$\dot P$ diagram shown in \autoref{fig:ppdot}. \autoref{fig:ppdot} also shows the population of known pulsars (gray dots) in the ATNF pulsar catalogue \citep{ATNF}, with binary pulsars indicated by circles. The black dashed line corresponds to the classical pulsar death line \citep{Chen+1993}. The black dash-dotted and dotted lines indicate alternative death-line models based on curvature radiation from vacuum gap (VG) and space-charged-limited flow (SCLF) scenarios \citep{Zhang+2000}. The gray dashed and gray dotted lines are lines of constant surface magnetic field and constant characteristic age, respectively \citep{hpa}.

Since these binaries contain low-mass ($\lesssim 1~M_\odot$) main-sequence companions with evolutionary timescales comparable to that of the Sun, their typical ages are estimated to be $\sim 10^{10}$~yr \citep[e.g.,][]{GaiaNS1,El-Badry+2024}. If any pulsars in these systems share a similar age, our evolutionary tracks and lines of constant characteristic age imply they should lie below the death line (highlighted in red) and thus have ceased emitting radio pulses. In this case, the nondetections could be explained by the pulsars having turned off (i.e., becoming radio-quiet). However, confirming whether pulsars below the death line can still produce radio emission requires a larger sample, in order to separate intrinsic emission shutoff from observational biases, such as beaming effects discussed in {Section}~\ref{subsec:prob}.

Most known binary pulsars with ages $\sim 10^{10}~{\rm yr}$ are recycled MSPs, as seen in the lower left region of \autoref{fig:ppdot}. In particular, compact binaries with very low-mass companions, e.g., eclipsing spider pulsars, are known to host active MSPs \citep{Roberts+2013}. Hot subdwarf-B (sdB) binaries have also been proposed as potential MSP hosts \citep{Podsiadlowski+2002,Geier+2010}. However, targeted radio searches of several sdB+NS candidates have so far resulted in nondetections \citep{Coenen+2011}. Nevertheless, the companions in these compact systems are often optically detectable, allowing the systems to be identified through optical surveys. Consequently, future radio pulsation searches targeting binaries with sdB or very low-mass companions may yield a higher detection rate than the binaries observed in this study and should be prioritized in forthcoming observational efforts.

{\subsection{Comparison with Known Binary Pulsars with Optical Companions}}

{While there have been a few case studies of searching for radio pulsations from NS candidates in detached binaries \citep{Yi+2022,Lin+2023,Zheng+2022,Brylyakova+2025}, none has yielded pulsar detections. These non-detections are primarily due to the radio beam not intersecting Earth or the NS being old and radio-quiet, as discussed above.}

{In contrast, substantial work has focused on the reverse approach: starting from known binary radio pulsars and searching for their optical companions using Gaia data \citep[e.g.,][]{Jennings+2018,Antoniadis+2020,Antoniadis+2021}. For example, by cross-matching 1534 known pulsars with sources in the second Gaia data release, \cite{Antoniadis+2021} identified 22 reliable associations, 18 of which are known MSPs. The majority of these binaries are eclipsing MSP binaries (so-called spider pulsars) with stellar/Jupiter-mass companions that are significantly bluer and more luminous than main-sequence stars of comparable masses (see Fig.~2 of \citealt{Antoniadis+2021}). This reflects mass transfer and ablation, where the companion has been stripped and heated by the pulsar wind. In contrast, the binaries investigated here are typically wide, detached systems where mass transfer has not yet begun. The NSs are therefore more likely to be normal (non-recycled) pulsars rather than MSPs, and their companions are ordinary main-sequence stars (see Fig.~1 of \citealt{El-Badry+2024}) without significant stripping or ablation.}

{Beyond spider systems, a few binary pulsars host luminous non-degenerate stellar companions. Notable examples include PSR~J0045--7319 \citep{Kaspi+1994}, in the Small Magellanic Cloud, with a B1V companion of mass $\sim 8.8~{M_\odot}$ in a 51-day orbit; PSR~B1259--63 \citep{Johnston+1992}, with an $\sim 8~{M_\odot}$ Be star companion in a highly eccentric 3.4-year orbit; PSR~J2032+4127 \citep{Lyne+2015}, orbiting a Be star in the Cyngus~OB2 association with a period of $\sim$ 30~yr; and PSR~J2108+4516 \citep{Andersen+2023}, with a massive OBe star in a 269~day orbit. These systems contain massive, young companions, in contrast to the low-mass companion of the NS candidates investigated here. Their pulsars remain radio-active, consistent with typical ages of $\lesssim 10^{7}~{\rm yr}$. Because the companions are much more massive than the NSs, measuring radial velocities and orbital separations is generally more challenging than for the systems in our targets.}

{Another exceptional case is PSR~J1903+0327 \citep{Champion+2008}, an MSP likely paired with a solar-type main-sequence companion. This system is distinct from the canonical MSP population and likely formed through dynamical interactions. However, PSR~J1903+0327 lies at a distance of $\sim6.4~{\rm kpc}$, which is much further than the NS candidates presented here. Combined with its 90-day orbital period, these two factors make it difficult for Gaia to obtain a robust astrometric solution.}

\section{summary}\label{sec:sum}

We conducted a deep radio pulsation search targeting 31 NS candidates in detached binary systems, using 46.65 hours of observations with FAST, GBT, and TMRT. No periodic or single-pulse emission was detected from any target.
These nondetections correspond to stringent upper limits on pulsed flux densities, down to a few $\mu \rm Jy$ for periodic signals (FAST) and $\sim 10~{\rm mJy}$ for single pulses (FAST), which places tight constraints on the radio luminosities of any underlying pulsars.
We estimated the detection probability using a log-normal luminosity function, incorporating the effects of geometric beaming.
The probability of missing all 31 sources is very low unless their emission beams are strongly misaligned with our line of sight, suggesting that geometric beaming is the most likely explanation, if these objects are indeed pulsars.
Alternatively, the NSs may be sufficiently old to have crossed the pulsar death line and become radio quiet. Our findings underscore the challenge of confirming NSs in old detached binary systems through radio pulsation searches.

\begin{acknowledgments}
\quad We are grateful to the referee for valuable comments that helped improve the manuscript. We thank Scott Ransom and Lei Zhang for helpful discussions on distinguishing astrophysical pulses in the \texttt{PRESTO}'s \texttt{prepfold} outputs. We thank Evan Smith for his kind assistance during observations with GBT as well as Wen-Ting He and Ying-Ao Tang for their assistance during in-site observations with TMRT. We thank Yi-Ying Wang for providing the beautiful sketch illustrating binary evolution.

This work made use of data from FAST (Five-hundred-meter Aperture Spherical radio Telescope)(\url{https://cstr.cn/31116.02.FAST}). FAST is a Chinese national mega-science facility, operated by National Astronomical Observatories, Chinese Academy of Sciences.
The Green Bank Observatory is a facility of the National Science Foundation operated under cooperative agreement by Associated Universities, Inc.

S.J.G. acknowledges support from the National Natural Science Foundation of China (NSFC) under grant No.~123B2045. 
X.D.L. acknowledges support from the National Key Research and Development Program of China (2021YFA0718500), the National Natural Science Foundation of China (NSFC) under grant No.~12041301, 12121003 and 12203051.
K.E. acknowledges support from NSF grant AST-2307232.
P.W. acknowledges support from the National Natural Science Foundation of China (NSFC) Programs No.~11988101, 12041303, the CAS Youth Interdisciplinary Team, the Youth Innovation Promotion Association CAS (id.~2021055), and the Cultivation Project for FAST Scientific Payoff and Research Achievement of CAMS-CAS.
P.Z. acknowledges support from the National Natural Science Foundation of China (NSFC) under grant No.~12273010.

The computation was made by using the facilities at the High-Performance Computing Center of Collaborative Innovation Center of Advanced Microstructures (Nanjing University).
\end{acknowledgments}

\section*{Data Availability}
FAST data are open {source} in the FAST Data Center according to the FAST data one-year protection policy. GBT data can be accessed by contacting the GBT staff with project ID: AGBT24A\_092. TMRT data will be shared on reasonable request to the corresponding author.

\facilities{FAST, GBT and TMRT.}

\software{\texttt{FETCH} \citep{FETCH},
\texttt{Heimdall} \citep{Barsdell+2024}, 
\texttt{PRESTO} \citep{Ransom+2011}, 
\texttt{PulsarX} \citep{PulsarX}, 
\texttt{Psrqpy} \citep{psrqpy},
\texttt{PyGEDM} \citep{pygedm}, 
\texttt{TransientX} \citep{TransientX} and 
\texttt{YOUR} \citep{Aggarwal+2020}.}

\appendix
\restartappendixnumbering

\section{Pulsar Search pipeline tests}\label{sec:app}
In this Appendix, we describe our search pipeline validation using GBT observation of PSR~J1503+2111 and outline the criteria by which we distinguish astrophysical pulses from RFI.

\subsection{Acceleration Search Candidates}

\begin{figure}
    \centering
    \includegraphics[width=\linewidth]{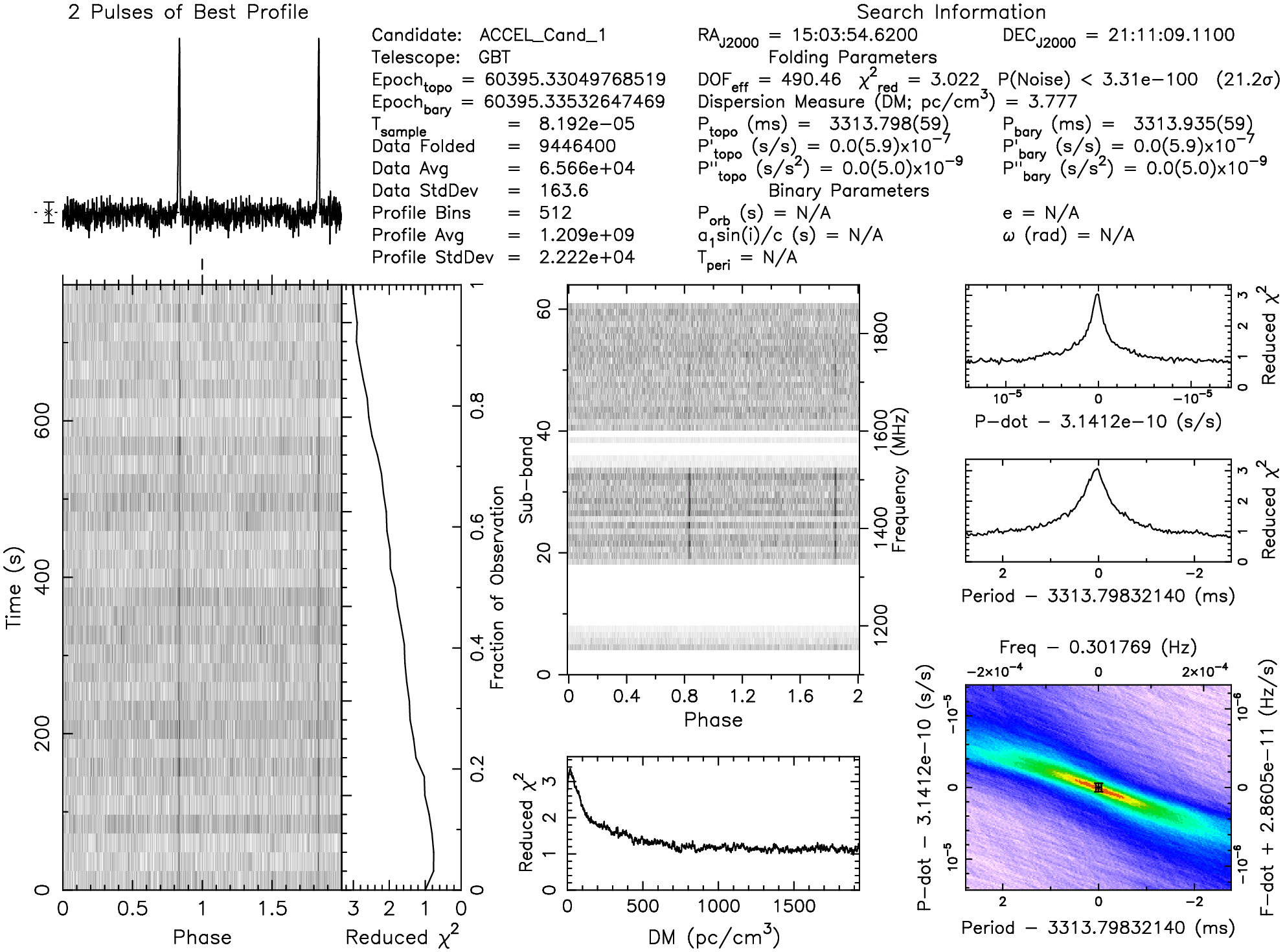}
    \caption{Diagnostic plot for PSR~J1503$-$2111. Upper left: Summed profile. Lower left: time--phase plot. Upper middle: frequency--phase plot. Lower middle: reduced $\chi^2$ vs. DM. Upper right: reduced $\chi^2$ vs. period derivative. Middle right: reduced $\chi^2$ vs. period. Lower right: reduced $\chi^2$ map in the period derivative and period plane. Observing and derived parameters are listed in the inset.}
    \label{fig:pfd_psr}
\end{figure}

\autoref{fig:pfd_psr} presents the folding results for PSR~J1503$-$2111, obtained with the \texttt{prepfold} routine in \texttt{PRESTO}. To distinguish candidates of astrophysical origin from those due to RFI, we examine at least four diagnostic features in the \texttt{prepfold} diagnostic plots \citep[for details, see][]{Zhu+2014}. Here we take \autoref{fig:pfd_psr} for example. (1) Summed profile (upper left panel). Real pulsar signals exhibit one or more narrow peaks. (2) Time--phase plot (lower left panel). This plot sums over all radio frequency channels. Real pulsar signals appears as one or more narrow vertical stripes aligned with the peaks in the summed profile. (3) Frequency--phase plot (upper middle panel). This plot sums over all observing time intervals. Real pulsar signals also appear as one or several narrow vertical stripes corresponding both to the peaks in the summed profiles and to the stripes in the time--phase plot. (4) Reduced $\chi^2$ vs. DM (lower middle panel). The $\chi^2$ is computed for each trial DM and higher reduced $\chi^2$ values indicate greater deviation from white noise. A real pulsar signal produces a clear peak at a nonzero DM.
All four criteria are met by the candidate shown in \autoref{fig:pfd_psr}, confirming its identification as PSR~J1503$-$2111. Although its DM is low, the reduced chi‐squared curve nevertheless peaks at a nonzero value.

\subsection{FFA Candidates}

\begin{figure}
    \centering
    \includegraphics[width=\linewidth]{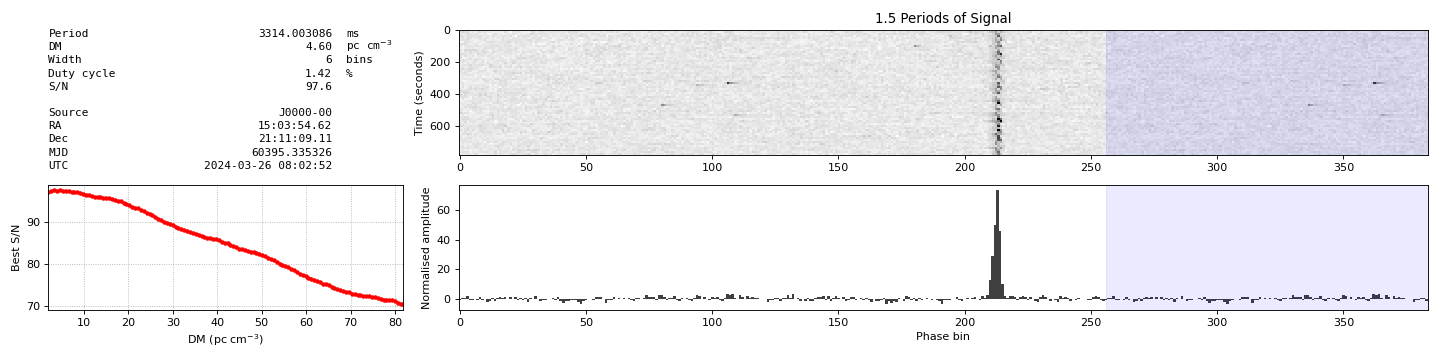}
    \caption{FFA diagnostic for PSR~J1503+2111. Upper left inset: observing parameters and detected signal properties. Lower left: S/N vs. DM. Upper right: time--phase plot. Lower right: summed pulse profile.}
    \label{fig:ffa_psr}
\end{figure}

\autoref{fig:ffa_psr} present the \texttt{rffa} output for PSR~J1503+2111. The inset in the upper left lists the observing parameters and the properties of the detected periodic signal. The classification criteria {are} similar to those used in \texttt{prepfold} diagnostics. The lower left panel plots the signal‐to‐noise ratio vs. DM, where a true pulsar produces a distinct peak at a nonzero DM. The upper right panel presents the time--phase plot, revealing a clear vertical stripe characteristic of dispersed pulses. The lower right panel displays the summed pulse profile, which exhibits a sharp peak at the folded period. Because \texttt{rffa} omits the frequency--phase plot, any promising candidate must be folded subsequently with \texttt{prepfold} and examined for narrow‐band RFI signatures to confirm its astrophysical origin.
The candidate shown in \autoref{fig:ffa_psr} is identified as PSR~J1503$-$2111 with a spin period of 3314.003086 ms and a DM of $4.6~{\rm pc~cm^{-3}}$, slightly higher than the value $3.26~{\rm pc~cm^{-3}}$ in the literature.

\subsection{Single-pulse Candidates}

\begin{figure}
    \centering
    \includegraphics[width=0.8\linewidth]{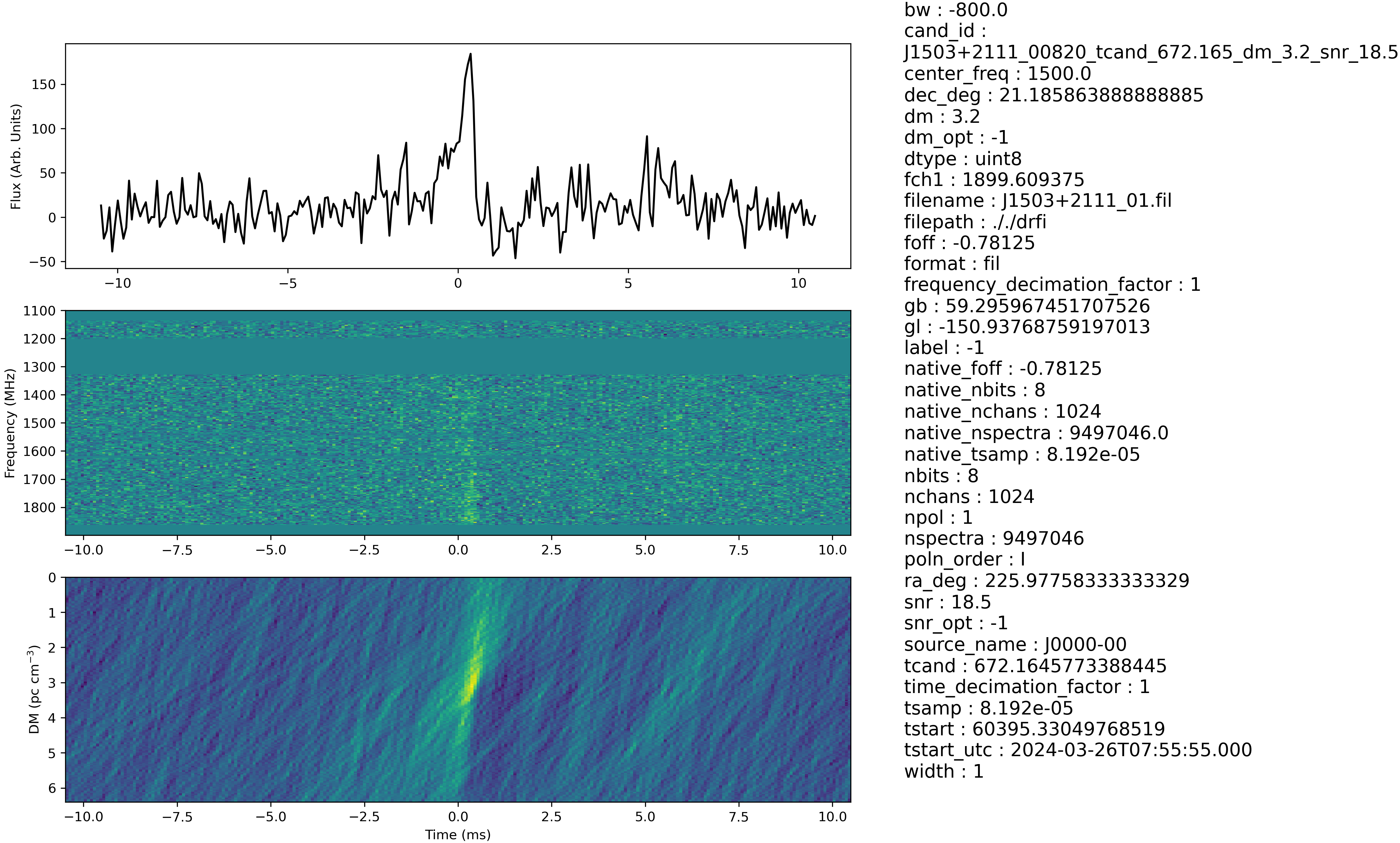}
    \caption{Single-pulse diagnostic for PSR~J1503+2111. Top: summed profile. Middle: dedispersed dynamic spectrum. Bottom: S/N map in time--DM plane.}
    \label{fig:sp_psr}
\end{figure}

The single pulse shown in \autoref{fig:sp_psr} was simultaneously detected by both \texttt{TransientX} and \texttt{Heimidall}, and subsequently identified by \texttt{FETCH} as an astrophysical event. In the top panel, a peak marks the pulse arrival. The dedispersed dynamical spectrum shown in the middle panel reveals a clear, vertical stripe across the band, indicating the dispersion sweep expected from a pulsar signal. In the bottom panel, the S/N map in the time--DM plane exhibits a clear maximum at DM$=3.2~{\rm pc~cm^{-3}}$, coinciding with the optimal arrival time. These features confirm that the candidate originates from PSR~J1503+2111 and is not due to RFI.

\bibliography{ref.bib}{}
\bibliographystyle{aasjournalv7}

\end{document}